\documentclass[11pt]{article}
\IfFileExists{jheppub.sty}
{\usepackage{jheppub}}
{\usepackage{jheppub_local}}
\usepackage{float}
\usepackage{amsmath,amssymb,amsfonts,amsthm}
\usepackage{booktabs}
\usepackage{multirow}
\usepackage{array}
\usepackage{xcolor}
\usepackage{mathrsfs}

\title{Anomaly-Free Spectra, Unimodular Lattices and 6D R-Symmetry Gauged Supergravity}

\author[a]{Katrin Becker and Qi You}

\affiliation[a]{%
George P. and Cynthia Woods Mitchell Institute for Fundamental Physics and Astronomy,
Texas A\&M University,
College Station, TX 77843, USA
}

\emailAdd{kbecker@physics.tamu.edu}
\emailAdd{qi\_you@tamu.edu}

\abstract{
We study the classification problem for anomaly-free 6D $\mathcal N=(1,0)$ supergravities with a gauged abelian R-symmetry and one tensor multiplet. We present eleven new models with gauge group $G_{\mathrm{non-Abelian}}\times U(1)_R$ that satisfy the local Green--Schwarz factorization condition, together with several recently proposed global consistency conditions. In particular, the low-rank models we found are precisely where some of the recent enumeration literature is least directly applicable. These examples suggest that the landscape of anomaly-free gauged $U(1)_R$ supergravities may be richer than previously recognized while still remaining highly constrained. We analyze the arithmetic structure of the anomaly coefficients, including their integral pairings, embeddability into rank-two unimodular charge lattices, the characteristic-vector condition and ghost-free gauge-field conditions. We show that $n_V \equiv 8 \pmod{12}$ is necessary and sufficient for the unimodular embeddability in the rank-two case, when the gauge group does not contain $SU(2)$, $SU(3)$ and $G_2$. For the characteristic-vector condition we verify sufficiency for the branches realized by our examples and identify a remaining branch requiring additional exclusion. We also present a detailed discussion of the contribution to the anomaly polynomial when the $D_4$ Lie algebra is present. These results sharpen the boundary between anomaly-free 6D spectra, global-consistency constraints, and possible UV realization in string theory or F-theory.}

\begin{document}

\maketitle

\tableofcontents

\section{Introduction}
\label{sec:introduction}

Six-dimensional $\mathcal{N}=(1,0)$ supergravity theories provide a highly constrained arena for studying the consistency conditions of quantum gravity. The cancellation of gravitational, gauge, and mixed anomalies imposes strong restrictions on the allowed gauge groups, matter representations, and Green--Schwarz couplings. These restrictions, together with additional global consistency conditions, make six dimensions especially well suited for systematic searches for consistent supergravity theories and for probing swampland constraints.

In recent years, many consistency conditions for six-dimensional $\mathcal{N}=(1,0)$ Poincaré supergravity theories and their R-symmetry gaugings have been proposed and refined.
We briefly summarize some of the most established consistency conditions, together with several others that we assume to hold and will play a central role in this work. For further discussion, see also~\cite{Becker:2023zyb, Monnier:2017oqd, Suzuki:2005vu, Monnier:2018nfs,Pang:2020rir}.
\begin{itemize}
    \item Local anomaly cancellation
\end{itemize}
Local anomaly cancellation requires that the anomaly polynomial (in general non-vanishing) 
must factorize and can therefore be canceled
at the quantum level via a tensor field coupling, which is known as the (generalized) Green--Schwarz mechanism \cite{Green:1984bx, Sagnotti:1992qw,Erler:1993zy}. 
When there is a single tensor multiplet, factorization reduces to the condition that the anomaly matrix has rank at most two \cite{Avramis:2005hc}.

\begin{itemize}
    \item Global anomaly cancellation
\end{itemize}
Global anomaly cancellation requires that the partition function be invariant under large gauge and diffeomorphism transformations, and that the Green--Schwarz counterterm be globally well-defined on general backgrounds. The latter is related to the Dai--Freed anomaly \cite{Dai:1994kq,Freed:2016rqq}. In the formulation of Monnier and Moore \cite{Monnier:2018cfa,Monnier:2018nfs}, global consistency is controlled by a combination of conditions: local anomaly factorization, unimodularity of the string charge lattice, integrality and parity constraints on the anomaly coefficients, and a residual topological obstruction characterized by spin bordism $\Omega_7^\mathrm{Spin}(BG)$. We will discuss the unimodularity and additional constraints on the anomaly coefficients in the next few paragraphs. As for the spin bordism condition, it has been proved that $\Omega_7^\mathrm{Spin}(BG)=0$ for products of $U(1)$ and any simply-connected simple groups \cite{Monnier:2018nfs,Becker:2025xgy}. For this reason, we only consider $U(1)$ and simply-connected  Lie groups in this paper.

\begin{itemize}
    \item String charge quantization
\end{itemize}
In analogy with Maxwell theory, where charge quantization follows from the consistency of Dirac monopole backgrounds, abelian two-form gauge fields in six dimensions require that the lattice of string charges form an integral unimodular lattice compatible with the self-duality condition of the tensor fields. In six-dimensional $\mathcal{N}=(1,0)$ supergravity, this charge lattice is identified with the lattice of Green–Schwarz coefficients, and its integrality is tightly constrained by anomaly cancellation conditions \cite{Kumar:2010ru,Seiberg:2011dr}.

\begin{itemize}
    \item Completeness hypothesis
\end{itemize}
It has been conjectured that in a consistent quantum gravity theory, all charges that are allowed by the Dirac quantization condition must be represented in the charge lattice \cite{Banks:2010zn}. A stronger formulation of this idea asserts that a consistent quantum gravity theory should admit a well-defined path integral on arbitrary spin manifolds, without requiring additional global structure beyond the spin structure itself. In this framework, any smooth gauge field configuration consistent with the quantization conditions is allowed in the functional integral, and global consistency conditions such as anomaly cancellation and bordism triviality must hold for all such backgrounds. This viewpoint naturally connects the completeness of the charge lattice with modern formulations of global anomalies in terms of invertible field theories and cobordism classification \cite{McNamara:2019rup}.

\begin{itemize}
    \item Additional conjectured constraints
\end{itemize}
In the context of F-theory compactifications to six dimensions, the data of the low-energy $\mathcal{N}=(1,0)$ supergravity theory is geometrically encoded in the structure of elliptically fibered Calabi–-Yau threefolds, in which tensor multiplets correspond to the Kähler moduli of the base. In this setting, anomaly coefficients are identified with effective divisors in the base, and the string charge lattice is realized by the second homology lattice endowed with its intersection form \cite{Kumar:2009ac,Kumar:2010ru}. In particular, the gravitational anomaly coefficient vector $a$, which appears in the Green–Schwarz factorization condition, is geometrically associated with the anti-canonical class of the base surface.

The classification of six-dimensional supergravity theories without gauged R-symmetry has been investigated in considerable detail. Although anomaly cancellation is highly restrictive, large families of anomaly-free spectra are known \cite{Hamada:2023zol,Hamada:2025vga}. The situation appears to be more selective for theories with gauged $U(1)_R$ symmetry. The known examples are comparatively sparse \cite{Suzuki:2005vu,Becker:2023zyb,Becker:2025xgy}, and the simultaneous requirements of Green--Schwarz factorization, charge quantization, and global consistency substantially reduce the space of candidate theories. This motivates a systematic search for new anomaly-free models with a gauged $U(1)_R$ factor.

In this paper we present new anomaly-free models with gauge group
\begin{equation}
\begin{aligned}
    G &= G_{\rm non\mbox{-}Abelian}\times U(1)_R  \\
      &= \prod_{i=1}^{n}G_i\times U(1)_R .
      \label{eqn:gauge-group}
\end{aligned}
\end{equation}
We restrict attention to models with one tensor multiplet, $n_T=1$. This is the case most directly connected to the pseudo-Lagrangian formulation reviewed in \cite{Becker:2023zyb}. In the conventions used there, the relevant bosonic terms take the form
\begin{equation}
\begin{aligned}
      e^{-1}\mathcal{L}
= &\frac{1}{4} R
- \frac{1}{12} e^{2\phi} H_{\mu\nu\rho}^{\alpha} H^{\mu\nu\rho}_\alpha
- \frac{1}{4} \partial_\mu L_\alpha \partial^\mu L^\alpha
- \frac{1}{4} L_\alpha c_i^{\alpha} \operatorname{Tr}_i \left( F_{\mu\nu} F^{\mu\nu} \right)\\
&- \frac{1}{2} \left( L_\alpha c_{n+1}^{\alpha} \right)^{-1}
- \frac{1}{16} \epsilon^{\mu_1 \cdots \mu_6}
B_{\mu_1 \mu_2} X_{\mu_3 \cdots \mu_6}^{1}.
\label{eqn:pseudo-lagra}
\end{aligned}
\end{equation}
Here $L_\alpha$ $(\alpha=0,1)$ parametrizes the scalar coset $SO(1,1)$, the constants $c_i^\alpha$ and $c_{n+1}^\alpha$ determine the gauge and $U(1)_R$ couplings, and $X^1$ is the four-form entering the modified tensor-field strength. As explained in \cite{Becker:2023zyb}, the pseudo-Lagrangian must be supplemented by the appropriate self-duality condition, and supersymmetry imposes corresponding restrictions on the allowed couplings.

For a fixed gauge group, the anomaly polynomial is determined by the chiral fermion spectrum. The hypermultiplet fermions are neutral under the gauged $U(1)_R$, while the remaining chiral fields contribute to the pure $U(1)_R$ and mixed anomalies in the standard way \cite{Bagger:1983tt,Avramis:2005hc}. We therefore specify the matter representations under the non-Abelian gauge factors and then verify that the resulting anomaly polynomial factorizes. As a first check, all models presented below satisfy the irreducible gravitational anomaly condition
\begin{equation}
    n_H-n_V+29 n_T = 273 .
\end{equation}

The main results of this paper are organized in the following way. In \autoref{sec:constraints}, we discuss the well-known Green--Schwarz factorization condition with some emphasis on $D_4$ algebra due to its additional independent fourth-order Casimir. We show that the congruence condition proposed in \cite{Becker:2025xgy} is necessary and sufficient for the unimodular embeddability, at least when the gauge group does not contain $SU(2)$, $SU(3)$ and $G_2$. We find that the string charge lattice can be classified by the number of vector multiplets and the anomaly coefficients associated with non-Abelian gauge groups. The characteristic-element condition is satisfied by all such lattices except one. We also analyze the gauge-field ghost-free condition in the end of this section. In \autoref{sec:new-models}, we present a collection of new anomaly-free spectra with gauged $U(1)_R$ symmetry that satisfy all consistency conditions considered in this paper. These examples include models with relatively large-rank classical groups, such as the new $E_8\times SU(18)\times U(1)_R$ model, and low-rank  models such as $SU(2)\times SU(3)\times SU(3)\times U(1)_R$, which scarcely appear in earlier searches.
Together, these spectra provide additional evidence that the landscape of anomaly-free gauged $U(1)_R$ supergravities is substantially richer than previously recognized, while still exhibiting remarkable arithmetic rigidity. In \autoref{sec:arithmetic-analysis}, we analyze the arithmetic structure for a particular family of models with gauge group $SU(2)^p\times SU(3)^q\times U(1)_R$. It turns out that the allowed spectra are heavily constrained by the integrality of anomaly coefficients and that the problem of solving Diophantine equations due to Green--Schwarz factorization can be reduced to solving an algebraic equation involving the number of vector multiplets. Finally, we discuss the implications of our results and give an outlook in \autoref{sec:discussion-and-outlook}.

\section{General constraints in six dimensions}
\label{sec:constraints}

\subsection{Factorization of the anomaly polynomial}

A six-dimensional $N=(1,0)$ supergravity theory contains one gravity multiplet together with $n_T$ tensor multiplets, $n_V$ vector multiplets, and $n_H$ hypermultiplets. 
We restrict attention to the case \(n_T=1\). From the perspective of anomaly cancellation, this is arguably the most interesting regime: theories with one tensor multiplet are considerably more constrained than their \(n_T>1\) counterparts, yet the corresponding space of consistent models is manifestly nonempty. The resulting balance between rigidity and nontriviality makes the \(n_T=1\) setting especially well suited for a systematic investigation of anomaly-free six-dimensional supergravities with gauged \(U(1)_R\) symmetry. Below we summarize our conventions, following \cite{Becker:2023zyb} closely.

For theories with gauge group 
\begin{equation} 
G_1\times \dots\times G_n \times U(1)_R,
\end{equation} 
where $G_i$, $i=1,\dots,n$ are semi-simple, 
 the total eight-form anomaly polynomial takes the form 
\begin{align}\label{ai}
I_8 ={}&
\big(\operatorname{tr} R^2\big)^2
+\frac{1}{6}\operatorname{tr} R^2\, \sum_{i=1}^{n} {\mathcal C}_i \operatorname{tr} F_i^2
+\frac{1}{6}(n_V-20)F^2\operatorname{tr} R^2
\nonumber\\[2mm]
&-\frac{2}{3}\,\sum_{i=1}^{n} {\mathcal B}_i
\big(\operatorname{tr} F_i^2\big)^2
-\frac{2}{3}(n_V+4)F^4
\nonumber\\[2mm]
&+4\,\sum_{i<j}^n{\mathcal C}_{ij}
\operatorname{tr} F_i^2\operatorname{tr} F_j^2
-4F^2\,\sum_{i=1}^{n} c_{i,A}\operatorname{tr} F_i^2 .
\end{align}

Here $F$ denotes the field strength associated with the $U(1)_R$ factor and $F_i$, $i=1,\dots,n$ the field strengths associated with the group factor $G_i$. 
Traces in arbitrary irreducible representations $R,S$ are reduced to traces in the fundamental representation, denoted by $\mathrm{tr}$,  using the group-theoretical coefficients $a_{i, R }$, $b_{i, R }$, $c_{i, R }$ and $d_{i, R }$ defined by 
\begin{align}
    \mathrm{Tr}_ R F_i^2&=c_{i, R }\, \mathrm{tr}F_i^2\\
    \mathrm{Tr}_ R F_i^4&=a_{i, R }\, \mathrm{tr}F_i^4+b_{i, R }\, (\mathrm{tr}F_i^2)^2+d_{i, R }\,\tilde{F_i^4}.~~~~\label{eqn:def-a-b-d}
\end{align}
Details about these coefficients can be found in \autoref{app:group-coeff}. 
Here $\widetilde{F}_i^{\,4}$ denotes the additional fourth-order invariant that exists only for the $D_4$ algebra (see \autoref{app:4th-D4}). Since $D_4$ is the unique Lie algebra possessing three independent quartic Casimir invariants, the coefficients $d_{i, R }$ vanish for all Lie algebras other than $D_4$. In many previous analyses of six-dimensional anomaly cancellation, these terms did not appear because the corresponding constructions either did not involve $D_4$ factors or were restricted to representations for which $d_{i, R }=0$. For more general representations of $D_4$, however, the contributions proportional to $d_{i, R }$ must be included explicitly. The relevant formulas and conventions are summarized in \autoref{app:4th-D4}. 
Using this notation the coefficients in Eq. \eqref{ai} become
\begin{equation}
\begin{split}
 {\mathcal{B}_i} &= b_{i,A} - \sum_ R  n_{i, R } b_{i, R } , \\
 {\mathcal{C}_i} &= c_{i,A} - \sum_ R  n_{i, R } c_{i, R } , \\
 {\mathcal{C}_{ij}} &= \sum_{RS} n_{ij,RS}c_{i, R }c_{j,S} ,\qquad  i \neq j .
\end{split}
\end{equation}
Here $A$ denotes the adjoint representation,$n_{i,R}$ is the number of hypermultiplets under representation $R$ of group $G_i$, and $n_{ij,RS}$ is that simultaneously under representation $R$ of $G_i$ and $S$ of $G_j~(i\neq j)$.

In Eq.~\eqref{ai} we have omitted the irreducible terms proportional to  $ \mathrm{tr} R^4$ and  $\mathrm{tr} F_i^4$, $i=1,\dots,n$, since these must vanish in any theory admitting Green–Schwarz factorization.
The irreducible gravitational anomaly cancels provided
\begin{equation}
 n_H - n_V = 244.
\label{grav-anomaly}
\end{equation}
The condition for the vanishing of the coefficients of the irreducible $\mathrm{tr} F_i^4$, and  $\mathrm{tr}\tilde F_i^4$, terms is
\begin{align}
     {\mathcal{A}_i}&= a_{i,A}-\sum_ R n_{i, R }a_{i, R }=0,\\
     {\mathcal{D}_i}&= \sum_ R n_{i, R }d_{i, R }=0.
\end{align}

Anomaly cancellation requires that the total polynomial factorizes in a Green--Schwarz form. For $n_T=1$ this can be written as
\begin{equation}\label{aii}
I_8 = \frac{1}{2} \Omega_{\alpha\beta}Y_4^\alpha Y_4^\beta,
\end{equation}
with
\begin{equation}
Y_4^\alpha =\frac{1}{2} a^\alpha \, \mathrm{tr} R^2 + \sum_{i=1}^{n+1} b^\alpha_i\left( \frac{2}{\lambda_i}\, \mathrm{tr} F_i^2\right) .
\end{equation}
$\Omega$ is a symmetric bilinear form with signature $(1,1)$ given by 
\begin{equation}
    \Omega_{\alpha\beta}=\begin{pmatrix}
    0 & 1  \\
    1 & 0  
    \end{pmatrix}, \label{eqn:metric}
\end{equation}
 and $\lambda_i$ are normalization factors taking the values listed in 
 \autoref{tab:lambda}.
\begin{table}[h]
    \centering
\begin{center}
    \begin{tabular}{|c|c|c|c|c|c|c|c|c|c|}
\hline
 & $A_n$ & $B_n$ & $C_n$ & $D_n$ & $E_6$ & $E_7$ & $E_8$ & $F_4$ & $G_2$\\
\hline
$\lambda$ & 1 & 2 & 1 & 2 & 6 & 12 & 60 & 6 & 2\\
\hline
    \end{tabular}
    \caption{Normalization factor $\lambda$ for different groups.}
    \label{tab:lambda}
\end{center}
\end{table}
Moreover we set $\lambda_{n+1}=1$.

Equating Eqs. (\ref{ai}) and (\ref{aii}) determines the anomaly coefficients
$a$ and $b_i$, through\footnote{Our conventions differ from that in \cite{Becker:2023zyb} by a negative sign in all these equations, which essentially depends on the chiralities of all fermions and two-forms one chooses in relevant multiplets.}
\begin{align}
a \cdot a &= 8,  \\ 
a \cdot b_i &= \frac{1}{6}\lambda_i  {\mathcal{C}_i}, \\
b_i \cdot b_i &= -\frac{1}{3}\lambda_i^2  {\mathcal{B}_i}, \\
b_i \cdot b_j &= \lambda_i \lambda_j  {\mathcal{C}_{ij}},\\
a \cdot \frac{b_{n+1}}{2} &= \frac{1}{12} (n_V - 20), \\
b_i \cdot \frac{b_{n+1}}{2} &= -\frac{\lambda_i c_{i,A}}{2},\\
\frac{b_{n+1}}{2}\cdot \frac{b_{n+1}}{2} &= -\frac{1}{12} (n_V + 4).
\end{align}
Note that the scale degree of freedom can be used to set 
\begin{equation} 
a=(-2,-2).
\end{equation}

\subsection{Unimodularity of the dyonic string charge lattice}
\label{sec:unimodularity}

The anomaly coefficients can be identified with charges of dyonic strings coupled to the self-dual tensor fields of the six-dimensional theory~\cite{Duff:1996cf, Seiberg:2011dr, Monnier:2017oqd}. Consistency of the resulting charge lattice, together with Dirac quantization for self-dual tensor fields, requires the lattice to be unimodular.

A necessary condition for the anomaly coefficients to be embedded into a unimodular dyonic string charge lattice is
\begin{equation}
n_V \equiv 8 \pmod{12} .\label{eqn:nv=8mod12}
\end{equation}
In our computational search for new anomaly-free six-dimensional supergravity models with gauged $U(1)_R$ symmetry, we imposed this congruence condition from the outset.
Empirically, we found that all candidate models satisfying the local anomaly-factorization conditions also admitted embeddings into rank-two unimodular lattices. This observation motivates the following question: for the class of $n_T=1$ theories considered here, to what extent does the condition $n_V\equiv 8\pmod{12}$ already guarantee unimodular embeddability of the anomaly coefficients?

To answer this question, we first need the integrality properties of the anomaly coefficients. See~\cite{Kumar:2010ru} for the validity of the Eqs.~\eqref{eqn:a-dot-a}--\eqref{eqn:bi-dot-bj} when there are no $SU(2)$, $SU(3)$ or $G_2$ factors in $G_\mathrm{non-Abelian}$. When there are such groups, the integrality associated with these groups is equivalent to the triviality of the sixth homotopy group, which requires additional model-by-model inspection. In addition, Eqs.~\eqref{eqn:a-dot-bn+1} and~\eqref{eqn:bn+1-dot-bn+1} are guaranteed by Eq.~\eqref{eqn:nv=8mod12}, and finally Eq.~\eqref{eqn:bi-dot-bn+1} can be derived using the values of $\lambda$ and $c_{i,A}$ respectively listed in \autoref{tab:lambda} and \autoref{tab:adjoint-cA-horizontal}.
\begin{align}
a \cdot a &= 8, \label{eqn:a-dot-a} \\ 
a \cdot b_i &= \frac{1}{6}\lambda_i  {\mathcal{C}_i}\in\mathbb{Z}, \label{eqn:a-dot-bi}\\
b_i \cdot b_i &= -\frac{1}{3}\lambda_i^2  {\mathcal{B}_i}\in\mathbb{Z},\label{eqn:bi-dot-bi} \\
b_i \cdot b_j &= \lambda_i \lambda_j  {\mathcal{C}_{ij}}\in\mathbb{Z},\label{eqn:bi-dot-bj} \\
a \cdot \frac{b_{n+1}}{2} &= \frac{1}{12} (n_V - 20)\in\mathbb{Z}, \label{eqn:a-dot-bn+1}\\
b_i \cdot \frac{b_{n+1}}{2} &= -\frac{\lambda_i c_{i,A}}{2}\in\mathbb{Z},\label{eqn:bi-dot-bn+1}\\
\frac{b_{n+1}}{2}\cdot \frac{b_{n+1}}{2} &= -\frac{1}{12} (n_V + 4)\in\mathbb{Z} \label{eqn:bn+1-dot-bn+1}.
\end{align}
 
\begin{table}[h]
\centering
\begin{tabular}{c|cccccccc}
\hline
$G$ 
& $SU(N)$ 
& $Sp(n)$ 
& $Spin(N)$ 
& $F_4$ 
& $E_6$ 
& $E_7$ 
& $E_8$ 
& $G_2$ \\
\hline
$c_{i,A}$ 
& $2N$ 
& $2n+2$ 
& $N-2$ 
& $3$ 
& $4$ 
& $3$ 
& $1$ 
& $4$ \\
\hline
\end{tabular}
\caption{Adjoint quadratic coefficients $c_{i,A}$.}
\label{tab:adjoint-cA-horizontal}
\end{table}

\noindent
{\bf Claim.}
For the class of $n_T=1$ theories considered here, with anomaly coefficients satisfying the integrality conditions Eqs.~\eqref{eqn:a-dot-a}--\eqref{eqn:bn+1-dot-bn+1}, the anomaly vectors admit an embedding into a rank-two unimodular lattice whenever
\begin{equation}
n_V \equiv 8 \pmod{12} .
\end{equation}

\noindent
{\bf Proof.} First note that Eq.~\eqref{eqn:nv=8mod12}, together with Eqs.~\eqref{eqn:a-dot-bn+1} and~\eqref{eqn:bn+1-dot-bn+1}, implies that
\begin{equation}
b_{n+1}=(2,-1-k)
\qquad {\rm or} \qquad
b_{n+1}=(-1-k,2),
\end{equation}
where $k$ is a non-negative integer satisfying
\begin{equation}
n_V=12k+8 .
\end{equation}
Using the $\mathbb Z_2$ symmetry exchanging the two components of all anomaly
coefficients simultaneously, we may choose
\begin{equation}
b_{n+1}=(-1-k,2).
\end{equation}

Throughout this proof we work in the basis where
\begin{equation}
a=(-2,-2),
\qquad
\Omega=
\left(
\begin{array}{cc}
0 & 1\\
1 & 0
\end{array}
\right).
\end{equation}

We now analyze the arithmetic structure of the vectors
\begin{equation}
b_i=(x_i,y_i).
\end{equation}

The integrality conditions
\begin{equation}
a\cdot b_i=-2(x_i+y_i)\in \mathbb Z,
\end{equation}
and
\begin{equation}
\frac{b_{n+1}}{2}\cdot b_i
=
x_i-\frac{1+k}{2}y_i
\in \mathbb Z
\end{equation}
imply that the components $x_i,y_i$ are rational. Indeed, defining
\begin{equation}
m_i:=a\cdot b_i,
\qquad
n_i:=\frac{b_{n+1}}{2}\cdot b_i,
\end{equation}
one finds
\begin{equation}
x_i=
\frac{4n_i-(k+1)m_i}{2(k+3)},
\qquad
y_i=
-\frac{m_i+2n_i}{k+3}.
\end{equation}

Since $k\ge 0$, one has $k+3\neq 0$, and therefore both coordinates are rational.
We may thus write
\begin{equation}
b_i^\alpha=\frac{u_i^\alpha}{v_i^\alpha},
\end{equation}
where $u_i^\alpha$ and $v_i^\alpha$ are coprime integers.

Using the conditions
\begin{equation}
a\cdot b_i=-\frac{2u_i^0v_i^1+2u_i^1v_i^0}{v_i^0v_i^1}\in \mathbb Z,
\qquad
b_i\cdot b_i =\frac{v_i^0v_i^1}{2u_i^0u_i^1} \in \mathbb Z,
\end{equation}
one obtains
\begin{equation}
v_i^0v_i^1
\;\Big|\;
2u_i^0v_i^1+2u_i^1v_i^0,
\end{equation}
and
\begin{equation}
v_i^0v_i^1
\;\Big|\;
2u_i^0u_i^1\ .
\end{equation}

We first show that neither denominator can contain an odd prime factor.
Suppose, for contradiction, that an odd prime number $l$ divides $v_i^0$.
Since $u_i^0$ and $v_i^0$ are coprime, one has
\begin{equation}
l\nmid u_i^0 .
\end{equation}

From
\begin{equation}
v_i^0v_i^1 \mid 2u_i^0u_i^1 ,
\end{equation}
and the fact that $l\neq 2$, it follows that
\begin{equation}
l\mid u_i^1 .
\end{equation}

Substituting this into
\begin{equation}
v_i^0v_i^1
\;\Big|\;
2u_i^0v_i^1+2u_i^1v_i^0,
\end{equation}
one finds that
\begin{equation}
l\mid v_i^1 .
\end{equation}

This contradicts the coprimality of $u_i^1$ and $v_i^1$.
Thus $v_i^0$ contains no odd prime factors.
Repeating the same argument for $v_i^1$ gives
\begin{equation}
v_i^0=2^p,
\qquad
v_i^1=2^q .
\end{equation}

Suppose first that both $p$ and $q$ are positive.
Since $u_i^0$ and $u_i^1$ are coprime to their denominators,
both numerators are odd.
One then has
\begin{equation}
b_i\cdot b_i
=
\frac{2u_i^0u_i^1}{2^{p+q}}
=
\frac{u_i^0u_i^1}{2^{p+q-1}} .
\end{equation}

Because $u_i^0u_i^1$ is odd, integrality of $b_i\cdot b_i$ requires
\begin{equation}
p+q\le 1 .
\end{equation}

Hence at least one of $p$ and $q$ must vanish.

Now suppose $q=0$.
Then
\begin{equation}
a\cdot b_i
=
-2\left(
\frac{u_i^0}{2^p}+u_i^1
\right)
=
-\frac{u_i^0}{2^{p-1}}-2u_i^1 .
\end{equation}

Since $u_i^0$ is odd whenever $p\ge 1$, integrality of $a\cdot b_i$ implies
\begin{equation}
p\le 1 .
\end{equation}

The same argument with $0\leftrightarrow 1$ shows that if $p=0$, then
\begin{equation}
q\le 1 .
\end{equation}

Therefore
\begin{equation}
p,q\in\{0,1\},
\qquad
p+q\le 1 .
\end{equation}

Thus each $b_i$ is of one of the following three types:
\begin{equation}
(b_i^0,b_i^1)\in \mathbb Z\times \mathbb Z,
\end{equation}
\begin{equation}
(b_i^0,b_i^1)\in
\frac12\mathbb Z_{\rm odd}\times \mathbb Z,
\end{equation}
or
\begin{equation}
(b_i^0,b_i^1)\in
\mathbb Z\times \frac12\mathbb Z_{\rm odd}.
\end{equation}

The remaining integrality condition
\begin{equation}\label{aix}
\frac{b_{n+1}}{2}\cdot b_i
=
b_i^0-\frac{1+k}{2}b_i^1
\in \mathbb Z
\end{equation}
further constraints these possibilities.

If
\begin{equation}
b_i^0\in \frac12\mathbb Z_{\rm odd},
\qquad
b_i^1\in \mathbb Z,
\end{equation}
then $k$ must be even and $b_i^1$ must be odd.

If instead
\begin{equation}
b_i^0\in \mathbb Z,
\qquad
b_i^1\in \frac12\mathbb Z_{\rm odd}, \label{eqn:b_i^0-in-Z-b_i^1-in-half-Z-odd}
\end{equation}
then one must have
\begin{equation}
k\equiv 3 \pmod 4 .
\end{equation}

We now construct the corresponding unimodular lattices.

\medskip

\noindent
{\bf Case 1: $k\equiv 1 \pmod 4$.}

\medskip

In this case all anomaly coefficients are integral and can therefore be embedded into
the even unimodular lattice $\Gamma^{1,1}$ with basis
\begin{equation}
e_1=(1,0),
\qquad
e_2=(0,1).
\end{equation}

The Gram matrix is
\begin{equation}
\left(
\begin{array}{cc}
0 & 1\\
1 & 0
\end{array}
\right),
\end{equation}
so the lattice is even and unimodular.

Moreover, it contains
\begin{equation}
a=(-2,-2),
\qquad
\frac{b_{n+1}}{2}=
\left(
-\frac{1+k}{2},
1
\right), 
\end{equation}
since $(1+k)/2\in\mathbb Z$.

\medskip

\noindent
{\bf Case 2: $k$ even.}

\medskip

In this case the possible half-integral vectors have the form
\begin{equation}
b_i^0\in \frac12\mathbb Z_{\rm odd},
\qquad
b_i^1\in \mathbb Z_{\rm odd}.
\end{equation}

If instead $b_i^0$ and $b_i^1$ are integral, Eq.~(\ref{aix}) implies
\begin{equation}
b_i^1\in 2\mathbb Z .
\end{equation}

Thus the full set of allowed vectors is
\begin{equation}
\{(x,y): x\in \mathbb Z,\ y\in 2\mathbb Z\}
\cup
\left\{
(x,y):
x\in \frac12\mathbb Z_{\rm odd},
\ y\in \mathbb Z_{\rm odd}
\right\}.
\end{equation}

All anomaly coefficients can then be embedded into the odd unimodular lattice
$I^{1,1}$.
A convenient basis is
\begin{equation}
e_1=
\left(
\frac12,
1
\right),
\qquad
e_2=
\left(
-\frac12,
1
\right).
\end{equation}

The Gram matrix is
\begin{equation}
\left(
\begin{array}{cc}
1 & 0\\
0 & -1
\end{array}
\right),
\end{equation}
so the lattice is odd and unimodular.

A general lattice vector takes the form
\begin{equation}
m e_1+n e_2
=
\left(
\frac{m-n}{2},
m+n
\right),
\end{equation}
which reproduces precisely the allowed vectors above.

\medskip

\noindent
{\bf Case 3: $k\equiv 3 \pmod 4$.}

\medskip

In this case the vectors have either the form
\begin{equation}
b_i^0\in \mathbb Z,
\qquad
b_i^1\in \frac12\mathbb Z_{\rm odd},
\end{equation}
or $b_i^0,~b_i^1\in \mathbb Z$. In the latter case, the anomaly coefficients again embed into $\Gamma^{1,1}$, while in the first case, pairwise integrality
\begin{equation}
b_i\cdot b_j\in \mathbb Z
\end{equation}
implies that all integer vectors have even first component and that all half-integral
vectors have first components of the same parity.

If the half-integral vectors have even first component, the anomaly coefficients embed
into the even unimodular lattice
\begin{equation}
L_{3,{\rm even}}
=
{\rm Span}_{\mathbb Z}
\left\{
(2,0),
\left(
0,
\frac12
\right)
\right\}
=
\left\{
\left(
2m,
\frac n2
\right)
:
m,n\in\mathbb Z
\right\}.
\end{equation}

The Gram matrix is
\begin{equation}
\left(
\begin{array}{cc}
0 & 1\\
1 & 0
\end{array}
\right),
\end{equation}
so this lattice is even and unimodular.

If instead the half-integral vectors have odd first component, they embed into the odd
unimodular lattice
\begin{equation}
L_{3,{\rm odd}}
=
{\rm Span}_{\mathbb Z}
\left\{
\left(
1,
\frac12
\right),
\left(
1,
-\frac12
\right)
\right\}.
\end{equation}

A general vector in this lattice takes the form
\begin{equation}
m
\left(
1,
\frac12
\right)
+
n
\left(
1,
-\frac12
\right)
=
\left(
m+n,
\frac{m-n}{2}
\right).
\end{equation}

Hence
\begin{equation}
L_{3,{\rm odd}}
=
\left\{
(x,y)\in \mathbb Z\times \frac12\mathbb Z
\ \big| \
x\equiv 2y \pmod{2}
\right\}.\label{eqn:L3-odd}
\end{equation}

The Gram matrix is
\begin{equation}
\left(
\begin{array}{cc}
1 & 0\\
0 & -1
\end{array}
\right),
\end{equation}
so this lattice is odd and unimodular.

Thus, in all cases, the anomaly coefficients can be embedded into a rank-two unimodular
charge lattice containing
\begin{equation}
a,
\qquad
b_i,
\qquad
\frac{b_{n+1}}{2}.
\end{equation}

This establishes sufficiency of the congruence condition $n_V\equiv 8\pmod{12}$ for unimodular embeddability in the absence of $SU(2)$, $SU(3)$ and $G_2$, within the present $n_T=1$ framework and standard anomaly normalization.

\subsection{Characteristic element condition}

The anomaly coefficient vector $a$ is required to be a characteristic element of the
string charge lattice $\Lambda_S$ \cite{Monnier:2017oqd}. Explicitly,
\begin{equation}
a \cdot x \equiv x \cdot x \pmod 2,
\qquad
\forall x \in \Lambda_S .
\end{equation}

The unimodularity analysis of the previous subsection shows that the anomaly coefficients
can be embedded into one of several rank-two unimodular lattices. We now examine the
characteristic-vector condition in each case.

First consider the standard even lattice $\Gamma^{1,1}$ with basis
\begin{equation}
e_1=(1,0),
\qquad
e_2=(0,1).
\end{equation}
For a general vector
\begin{equation}
x=me_1+ne_2=(m,n),
\qquad
m,n\in\mathbb Z,
\end{equation}
one has
\begin{equation}
a\cdot x-x\cdot x
=
-2(m+n)-2mn
\in 2\mathbb Z .
\end{equation}
Thus $a$ is characteristic for $\Gamma^{1,1}$.

Next consider the odd unimodular lattice with basis
\begin{equation}
e_1=\left(\frac12,1\right),
\qquad
e_2=\left(-\frac12,1\right).
\end{equation}
A general vector is
\begin{equation}
x=me_1+ne_2
=
\left(
\frac{m-n}{2},
m+n
\right),
\qquad
m,n\in\mathbb Z .
\end{equation}
Then
\begin{equation}
a\cdot x-x\cdot x
=
-3m-n-m^2+n^2
=
-m(m+3)+n(n-1)
\in 2\mathbb Z .
\end{equation}
Hence $a$ is also characteristic for this odd unimodular lattice.
The same computation also applies to the lattice $L_{3,{\rm odd}}$ in Eq.~\eqref{eqn:L3-odd},
since it is obtained from this odd lattice by exchanging the two coordinates,
while $a=(-2,-2)$ is invariant under this exchange.

There remains, however, one additional lattice realization from Case 3 of the previous
subsection, namely
\begin{equation}
L_{3,{\rm even}}
=
{\rm Span}_{\mathbb Z}
\left\{
(2,0),
\left(
0,
\frac12
\right)
\right\}.
\end{equation}
For this lattice the characteristic condition is not automatic. Indeed, taking
\begin{equation}
x=
\left(
0,
\frac12
\right)
\end{equation}
gives
\begin{equation}
x\cdot x=0,
\qquad
a\cdot x=-1,
\end{equation}
so that
\begin{equation}
a\cdot x
\not\equiv
x\cdot x
\pmod 2 .
\end{equation}
Thus $a$ is not characteristic for the $L_{3,{\rm even}}$ branch.
Consequently, unimodular embeddability alone does not imply the
characteristic-vector condition. However, in the concrete search for anomaly-free models we never found a case with an $L_{3,{\rm even}}$ lattice, and there is a possibility that it can be excluded using representation theory analysis at the spectra level.

\subsection{Gauge-field ghost-free condition}

Aside from anomaly cancellation, a consistent supergravity theory must have gauge
kinetic terms with the correct sign. This requires the existence of a scalar-vector $L^\alpha$ such that
\begin{equation}
L_\alpha b_i^\alpha>0,\qquad i=1,\dots, n+1,
\label{2.116}
\end{equation}
for every gauge factor,which can be seen from the pseudo-Lagragian \eqref{eqn:pseudo-lagra}, using the fact that the gauge coupling $c_i$ is proportional to the anomaly-vector $b_i$ via a positive constant \cite{Becker:2023zyb}. In addition, the coefficient of the curvature-squared term must have the appropriate sign \cite{Kim:2019vuc}, which in our conventions requires
\begin{equation}
L_\alpha a^\alpha<0.
\label{2.117}
\end{equation}
The reason is that the gravitational anomaly vector $a^\alpha$ enters the Green-Schwarz $4$-form $Y_4^\alpha$ through the curvature term $\frac{1}{2} a^\alpha \text{tr}R^2$. The supersymmetric completion of this term generates higher-derivative curvature couplings proportional to $L_\alpha a^\alpha \text{tr}R^2$. To guarantee the absence of ghost modes in the gravitational sector, the overall coupling must satisfy \eqref{2.117}.

For $n_T=1$ we may parametrize the scalar as
\begin{equation}
L^\alpha=(e^\phi,e^{-\phi}) .
\label{2.118}
\end{equation}
Using
\begin{equation}
a=(-2,-2),
\qquad
b_{n+1}=(-k-1,2),
\label{2.119}
\end{equation}
the conditions become that there exists some $\phi\in\mathbb{R}$ that satisfies
\begin{equation}
b_i^0 e^\phi+b_i^1 e^{-\phi}>0,
\label{2.120}
\end{equation}
\begin{equation}
-(k+1)e^\phi+2e^{-\phi}>0,
\label{2.121}
\end{equation}
and
\begin{equation}
-2e^\phi-2e^{-\phi}<0 .
\label{2.122}
\end{equation}
The last inequality is automatically satisfied, so we focus on the first two. For convenience, we define
\begin{equation}
x=e^{2\phi}>0.
\label{2.123}
\end{equation}
If $b_i^0,~b_i^1>0$, the first inequality obviously holds and the second is solved by
\begin{equation}
0<x<\frac{2}{k+1},
\label{2.124}
\end{equation}
which is non-empty for arbitrary $k \geq 0$.

\noindent If $b_i^0,~b_i^1<0$, there is no solution for Eq.~\eqref{2.120}. The concrete models found in this work are not of this type. 

\noindent If $b_i^0>0,~b_i^1<0$, the solution set is non-empty as long as
\begin{equation}
0<-\frac{b_i^1}{b_i^0}<\frac{2}{k+1}. 
\end{equation}
If $b_i^0<0,~b_i^1>0$, the solution set is
\begin{equation}
0<x<\min\{-\frac{b_i^1}{b_i^0},~\frac{2}{k+1}\},
\end{equation}
which is again non-empty for $\forall k \geq 0$. 

\noindent The remaining two special cases are: 
if $b_i^0=0$, the condition reduces simply to $b_i^1>0$, and similarly if $b_i^1=0$, it reduces to $b_i^0>0$.

The anomaly coefficients of the models in \autoref{sec:new-models} are such that one of the above conditions with non-empty solution set can always be satisfied. Hence all models presented there have a region in tensor-multiplet moduli space in which the gauge kinetic terms and the curvature-squared term have the correct sign.

\section{New anomaly-free models}
\label{sec:new-models}

In this section we present eleven new candidate anomaly-free six-dimensional
$\mathcal{N}=(1,0)$ supergravity theories with gauged $U(1)_R$ symmetry.
All models satisfy the irreducible gravitational anomaly condition
\begin{equation}
n_H-n_V+29n_T=273,
\label{3.1}
\end{equation}
with $n_T=1$, admit Green--Schwarz factorization compatible with the
structure discussed in \autoref{sec:constraints}, and satisfy all other global consistency conditions mentioned in the Introduction.

For the model-searching process to be computationally tractable, we restrict the scope of our scan such that:
\begin{enumerate}
\item the Lie groups we consider here include $SU(N)~(2\leq N\leq18)$, $Spin(N)~(6\leq N\leq15)$, $Sp(N)~(2\leq N\leq10)$, $F_4$, $G_2$, $E_6$, $E_7$ and $E_8$;
\item the dimension of irreducible representions are smaller than $1000$;
\item the multiplicities of each product representations are set to be smaller than or equal to one.
\end{enumerate}
Despite these restrictions, we do not claim to provide an exhaustive list of spectra. Indeed, the Mathematica computations occasionally timeout or become intractable, particularly for gauge groups involving $SU(2)$ and $SU(3)$ due to their abundance of low-dimensional representations. Consequently, the models presented below should be viewed as an extension--rather than a complete classification--of 6D $\mathcal{N}=(1,0)$ supergravity theories with gauged $U(1)_R$ symmetry.

Rather than presenting each model separately through repetitive anomaly checks,
we organize the spectra according to their structural properties. The complete
matter content\footnote{Note that a factor of one-half may appear in the representation of matter content if it is pseudo-real.} and field content are summarized in \autoref{table:models},
while the corresponding Green--Schwarz anomaly coefficients are collected in
\autoref{table:factorization}. Several broad features emerge:

\begin{enumerate}
\item the existence of models with exceptional gauge factors;
\item the appearance of large-rank classical groups;
\item the persistence of unimodular charge lattices;
\item the strong arithmetic rigidity exhibited by the $SU(2)^p\times SU(3)^q\times U(1)_R$ models.
\end{enumerate}

The diversity of these spectra suggests that anomaly-free gauged
$U(1)_R$ supergravities are considerably richer than previously known,
while still occupying a highly constrained region of theory space.

\subsection{Exceptional-group and large-rank models}

Several of the new models involve exceptional gauge factors or unusually
large-rank classical groups.

Model I,
\[
E_7\times E_8\times Sp(2)\times U(1)_R,
\]
contains simultaneous exceptional and symplectic gauge sectors together with
matter transforming in higher-dimensional representations. This model admits an odd lattice embedding.

Model II,
\[
SU(6)\times Sp(2)\times F_4\times E_6\times U(1)_R,
\]
provides an especially interesting example containing two exceptional factors.
The coexistence of $F_4$ and $E_6$ with nontrivial matter couplings imposes
strong restrictions on the quartic and mixed gauge anomalies. This model admits an odd lattice embedding as well.

Model III,
\[
E_8\times SU(18)\times U(1)_R,
\]
contains a remarkably large-rank classical factor and can be embedded into an even lattice. The existence of such a
solution suggests that anomaly cancellation does not merely constrain the rank
of the gauge group, but rather imposes highly nontrivial arithmetic restrictions
on the allowed matter spectra.

\subsection{Classical-group models}

The remaining models exhibit a variety of classical gauge-group structures.
Several contain higher-dimensional or pseudo-real representations whose anomaly contributions combine in highly constrained ways. Among them, models IV--VII admit an even lattice embedding while models VIII and IX admit an odd lattice embedding.

Models IV and V,
\[
SU(2)\times Spin(10)\times Spin(14)\times U(1)_R,
\]
and
\[
SU(2)\times Spin(15)\times Sp(5)\times U(1)_R,
\]
illustrate the interplay between orthogonal and symplectic gauge factors.
The matter spectra involve large representations such as the ${\bf 364}$ of
$Spin(14)$, the ${\bf 128}$ of $Spin(15)$, and the ${\bf 110}$ of $Sp(5)$.

Models VI and VII are particularly noteworthy because they share the same gauge
group,
\[
SU(3)\times SU(6)\times U(1)_R,
\]
while possessing distinct matter spectra. This demonstrates that the anomaly
constraints can admit multiple isolated solutions for a fixed gauge group.
The existence of these distinct branches already hints at the arithmetic
structure analyzed systematically in \autoref{sec:arithmetic-analysis}.

Models VIII and IX,
\[
SU(2)\times Spin(7)\times Sp(5)\times U(1)_R,
\]
and
\[
Sp(2)\times Sp(9)\times Spin(12)\times U(1)_R,
\]
contain several pseudo-real representations and again exhibit highly
nontrivial factorization properties.

Taken together, these models suggest that the space of anomaly-free spectra is
highly structured rather than randomly distributed. In particular, the repeated
appearance of pseudo-real representations and highly constrained mixed sectors
appears closely tied to the arithmetic restrictions arising from anomaly
factorization.

\subsection{Low-rank models}

Models X and XI,
\[
SU(2)\times SU(3)\times SU(3)\times U(1)_R,
\]
admit an even lattice embedding and will play a distinguished role in the present work because they arise naturally from
the arithmetic analysis developed in \autoref{sec:arithmetic-analysis}. In addition, these models and the following arithmetic analysis complement earlier searches for $U(1)_R$ gauged models and classification/enumeration work such as \cite{Hamada:2024oap}.

These models occur in the first nontrivial branch satisfying the congruence
conditions implied by unimodularity and anomaly factorization. Their existence
provides strong evidence that the arithmetic constraints derived in Section~4
capture genuine structural features of the space of anomaly-free theories.

As explained in \autoref{sec:arithmetic-analysis}, the mixed anomaly equations eliminate the overwhelming
majority of naively allowed arithmetic branches. The surviving spectra therefore
appear to occupy isolated arithmetic islands within the space of anomaly-free
models.

\begin{table}[t]
\centering
\scriptsize
\renewcommand{\arraystretch}{1.2}
\begin{tabular}{ccccc}
\toprule
Model & Gauge group & Matter content & $n_H$ & $n_V$ 
\\
\midrule

I &
$E_7\times E_8\times Sp(2)\times U(1)_R$
&
$\frac12(\mathbf{1},\mathbf{248},\mathbf{4})\oplus\frac12(\mathbf{56},\mathbf{1},\mathbf{5})$
&
636 & 392
\\[6pt]

II &
$SU(6)\times Sp(2)\times F_4\times E_6\times U(1)_R$
&
$\begin{aligned}
&\frac12(\mathbf{1},\mathbf{4},\mathbf{1},\mathbf{78})\oplus\frac12(\mathbf{1},\mathbf{4},\mathbf{26},\mathbf{1})\oplus(\mathbf{6},\mathbf{1},\mathbf{1},\mathbf{27})\oplus\frac12(\mathbf{20},\mathbf{5},\mathbf{1},\mathbf{1})
\end{aligned}$
&
420 & 176
\\[6pt]

III &
$E_8\times SU(18)\times U(1)_R$
&
$(\mathbf{1},\mathbf{816})$
&
816 & 572
\\[6pt]

IV &
$SU(2)\times Spin(10)\times Spin(14)\times U(1)_R$
&
$\frac12(\mathbf{2},\mathbf{1},\mathbf{364})\oplus\frac12(\mathbf{4},\mathbf{10},\mathbf{1})$
&
384 & 140
\\[6pt]

V &
$SU(2)\times Spin(15)\times Sp(5)\times U(1)_R$
&
$\begin{aligned}
&\frac12(\mathbf{1},\mathbf{1},\mathbf{110})\oplus\frac12(\mathbf{2},\mathbf{128},\mathbf{1})\oplus\frac12(\mathbf{3},\mathbf{15},\mathbf{10})
\end{aligned}$
&
408 & 164
\\[6pt]

VI &
$SU(3)\times SU(6)\times U(1)_R$
&
$(\mathbf{3},\mathbf{21})\oplus(\mathbf{15},\mathbf{15})$
&
288 & 44
\\[6pt]

VII &
$SU(3)\times SU(6)\times U(1)_R$
&
$\frac12(\mathbf{1},\mathbf{540})\oplus(\mathbf{3},\mathbf{6})$
&
288 & 44
\\[6pt]

VIII &
$SU(2)\times Spin(7)\times Sp(5)\times U(1)_R$
&
$\begin{aligned}
&\frac12(\mathbf{1},\mathbf{1},\mathbf{220})\oplus\frac12(\mathbf{3},\mathbf{1},\mathbf{132})\oplus\frac12(\mathbf{4},\mathbf{8},\mathbf{1})
\end{aligned}$
&
324 & 80
\\[6pt]

IX &
$Sp(2)\times Sp(9)\times Spin(12)\times U(1)_R$
&
$\begin{aligned}
&\frac12(\mathbf{1},\mathbf{18},\mathbf{12})\oplus\frac12(\mathbf{4},\mathbf{152},\mathbf{1})\oplus\frac12(\mathbf{5},\mathbf{1},\mathbf{32})
\end{aligned}$
&
 492 & 248
\\[6pt]

X &
$SU(2)\times SU(3)\times SU(3)\times U(1)_R$
&
$\begin{aligned}
&(\mathbf{1},\mathbf{8},\mathbf{10})\oplus(\mathbf{1},\mathbf{10},\mathbf{8})\oplus(\mathbf{2},\mathbf{1},\mathbf{10})\oplus(\mathbf{2},\mathbf{10},\mathbf{1})\oplus\frac12(\mathbf{2},\mathbf{8},\mathbf{8})
\end{aligned}$
&
264 & 20
\\[6pt]

XI &
$SU(2)\times SU(3)\times SU(3)\times U(1)_R$
&
$\begin{aligned}
&(\mathbf{2},\mathbf{1},\mathbf{10})\oplus(\mathbf{2},\mathbf{10},\mathbf{1})\oplus3\cdot\frac12(\mathbf{2},\mathbf{8},\mathbf{8})\oplus\frac12(\mathbf{4},\mathbf{1},\mathbf{8})\oplus\frac12(\mathbf{4},\mathbf{8},\mathbf{1})
\end{aligned}$
&
264 & 20
\\[6pt]
\bottomrule
\end{tabular}
\caption{Matter spectra for the eleven anomaly-free models.}
\label{table:models}
\end{table}

\begin{table}[t]
\centering
\scriptsize
\renewcommand{\arraystretch}{1.25}
\begin{tabular}{clcl}
\toprule
Model & Green--Schwarz anomaly coefficients
& Model & Green--Schwarz anomaly coefficients \\
\midrule

I &
$\begin{array}{l}
b_{E_7}=(-\frac32,1),\;
b_{E_8}=(3,2),\\
b_{Sp(2)}=(\frac{27}{2},1),\;
b_{U(1)_R}=(-33,2)
\end{array}$
&
VII &
$\begin{array}{l}
b_{SU(3)}=(-1,1),\;
b_{SU(6)}=(8,7),\\
b_{U(1)_R}=(-4,2)
\end{array}$
\\[6pt]

II &
$\begin{array}{l}
b_{SU(6)}=(\frac32,1),\;
b_{Sp(2)}=(\frac92,1),\\
b_{F_4}=(-\frac32,1),\;
b_{E_6}=(3,2),\\
b_{U(1)_R}=(-15,2)
\end{array}$
&
VIII &
$\begin{array}{l}
b_{SU(2)}=(19,6),\;
b_{Spin(7)}=(-\frac32,1),\\
b_{Sp(5)}=(\frac92,3),\;
b_{U(1)_R}=(-7,2)
\end{array}$
\\[6pt]

III &
$\begin{array}{l}
b_{E_8}=(-6,1),\;
b_{SU(18)}=(6,1),\\
b_{U(1)_R}=(-48,2)
\end{array}$
&
IX &
$\begin{array}{l}
b_{Sp(2)}=(\frac{15}{2},1),\;
b_{Sp(9)}=(\frac12,1),\\
b_{Spin(12)}=(\frac12,1),\;
b_{U(1)_R}=(-21,2)
\end{array}$
\\[6pt]

IV &
$\begin{array}{l}
b_{SU(2)}=(16,3),\;
b_{Spin(10)}=(-2,1),\\
b_{Spin(14)}=(6,3),\;
b_{U(1)_R}=(-12,2)
\end{array}$
&
X &
$\begin{array}{l}
b_{SU(2)}=(1,3),\;
b_{SU(3)_1}=(9,12),\\
b_{SU(3)_2}=(9,12),\;
b_{U(1)_R}=(-2,2)
\end{array}$
\\[6pt]

V &
$\begin{array}{l}
b_{SU(2)}=(26,4),\;
b_{Spin(15)}=(1,2),\\
b_{Sp(5)}=(1,1),\;
b_{U(1)_R}=(-14,2)
\end{array}$
&
XI &
$\begin{array}{l}
b_{SU(2)}=(7,9),\;
b_{SU(3)_1}=(6,9),\\
b_{SU(3)_2}=(6,9),\;
b_{U(1)_R}=(-2,2)
\end{array}$
\\[6pt]

VI &
$\begin{array}{l}
b_{SU(3)}=(29,16),\;
b_{SU(6)}=(2,4),\\
b_{U(1)_R}=(-4,2)
\end{array}$
\\[6pt]

\bottomrule
\end{tabular}
\caption{Green--Schwarz anomaly coefficients for the eleven anomaly-free models.}
\label{table:factorization}
\end{table}

\section{Structure of the search for \texorpdfstring{$SU(2)^p \times SU(3)^q \times U(1)_R$}{SU(2)p x SU(3)q x U(1)R} models}
\label{sec:arithmetic-analysis}

In this section we analyze the arithmetic structure underlying anomaly-free six-dimensional
$N=(1,0)$ supergravity theories with gauge group
\begin{equation}
G = SU(2)^p \times SU(3)^q \times U(1)_R,
\qquad n_T = 1 .
\label{5.1}
\end{equation}
Our goal is to understand how the local anomaly equations, together with the global unimodularity constraint, restrict the space of admissible spectra.

We will see that the anomaly equations impose strong arithmetic divisibility and positivity constraints, leading to a remarkably sparse set of admissible spectra. After imposing the global congruence condition and solving the Green--Schwarz equations, the classification problem reduces to a highly constrained arithmetic search controlled by a small number of integer parameters. The combined quadratic, mixed, and integrality constraints leave only a very sparse set of admissible branches.

\subsection{Gravitational anomaly constraints}

For the gauge group
\begin{equation}
G = SU(2)^p \times SU(3)^q \times U(1)_R,
\label{5.2}
\end{equation}
the total number of vector multiplets is
\begin{equation}
n_V = 1 + 3p + 8q
\label{5.3}
\end{equation}

The unimodularity condition derived in \autoref{sec:constraints} requires
\begin{equation}
n_V \equiv 8 \pmod{12}.
\label{5.4}
\end{equation}
Substituting Eq.~(\ref{5.3}) gives
\begin{equation}
3p + 8q + 1 \equiv 8 \pmod{12},
\label{5.5}
\end{equation}
or equivalently
\begin{equation}
3p + 8q \equiv 7 \pmod{12}.
\label{5.6}
\end{equation}

Reducing modulo $4$ immediately yields
\begin{equation}
3p \equiv 3 \pmod{4},
\label{5.7}
\end{equation}
which implies
\begin{equation}
p \equiv 1 \pmod{4}.
\label{5.8}
\end{equation}
Similarly, reducing modulo $3$ gives
\begin{equation}
2q \equiv 1 \pmod{3},
\label{5.9}
\end{equation}
and therefore
\begin{equation}
q \equiv 2 \pmod{3}.
\label{5.10}
\end{equation}

Hence we may write
\begin{equation}
p = 4r + 1,
\qquad
q = 3\ell + 2,
\label{5.11}
\end{equation}
with $r,\ell \in {\mathbb Z}_{\geq 0}$.

Substituting these expressions into Eq.~(\ref{5.3}) gives
\begin{equation}
n_V
=
1 + 3(4r + 1) + 8(3\ell + 2)
=
20 + 12r + 24\ell .
\label{5.12}
\end{equation}

It is convenient to define
\begin{equation}
k := r + 2\ell + 1,
\label{5.13}
\end{equation}
so that
\begin{equation}
n_V = 8 + 12k .
\label{5.14}
\end{equation}

Using the gravitational anomaly condition
\begin{equation}
n_H - n_V + 29n_T = 273,
\label{5.15}
\end{equation}
with $n_T = 1$, one obtains
\begin{equation}
n_H = 244 + n_V = 252 + 12k .
\label{5.16}
\end{equation}

Thus, for each fixed positive integer $k$, the gravitational anomaly constraints reduce the classification problem to a finite set of possible $(p,q)$ choices.

\subsection{Local anomaly structure}

We now analyze the Green--Schwarz factorization conditions.

Throughout this section we work in the basis
\begin{equation}
a = (-2,-2),
\qquad
\Omega =
\begin{pmatrix}
0 & 1 \\
1 & 0
\end{pmatrix},
\label{5.17}
\end{equation}
and write
\begin{equation}
b_{n+1} = (-k-1,2).
\label{5.18}
\end{equation}

For each simple gauge-group factor $G_i$ we denote the corresponding anomaly coefficient by
\begin{equation}
b_i = (x_i,y_i).
\label{5.19}
\end{equation}

The anomaly equations are
\begin{equation}
a \cdot b_i =
\frac{1}{6}\lambda_i {\mathcal{C}_i},
\label{5.20}
\end{equation}
\begin{equation}
b_i \cdot \frac{b_{n+1}}{2}
=
-\frac{1}{2}\lambda_i c_{i,A},
\label{5.21}
\end{equation}
and
\begin{equation}
b_i \cdot b_i
=
-\frac{1}{3}\lambda_i^2 {\mathcal{B}_i} .
\label{5.22}
\end{equation}

Since
\begin{equation}
a \cdot b_i = -2x_i - 2y_i,
\label{5.23}
\end{equation}
and
\begin{equation}
b_i \cdot b_{n+1}
=
2x_i - (k+1)y_i,
\label{5.24}
\end{equation}
Eqs.~(\ref{5.20}) and (\ref{5.21}) become
\begin{equation}
-2x_i -2y_i =
\frac{1}{6}\lambda_i{\mathcal{C}_i},
\label{5.25}
\end{equation}
\begin{equation}
2x_i -(k+1)y_i =
-\lambda_i c_{i,A} .
\label{5.26}
\end{equation}

Solving for $(x_i,y_i)$ gives
\begin{equation}
x_i =
-\frac{
\lambda_i\big((k+1){\mathcal{C}_i} + 12c_{i,A}\big)
}{
12(k+3)
},
\label{5.27}
\end{equation}
\begin{equation}
y_i =
\frac{
\lambda_i(6c_{i,A} -{\mathcal{C}_i})
}{
6(k+3)
}.
\label{5.28}
\end{equation}

Hence
\begin{equation}
b_i =
\left(
-\frac{
\lambda_i\big((k+1){\mathcal{C}_i} + 12c_{i,A}\big)
}{
12(k+3)
},
\,
\frac{
\lambda_i(6c_{i,A} -{\mathcal{C}_i})
}{
6(k+3)
}
\right).
\label{5.29}
\end{equation}

Substituting Eq.~(\ref{5.29}) into Eq.~(\ref{5.22}) yields
\begin{equation}
(6c_{i,A} - {\mathcal{C}_i})
\big((k+1){\mathcal{C}_i} + 12c_{i,A}\big)
=
12(k+3)^2{\mathcal{B}_i} .
\label{5.30}
\end{equation}

This relation is the key arithmetic constraint governing the allowed matter spectra.

\subsection{Specialization to \texorpdfstring{$SU(2)^p \times SU(3)^q$}{SU(2)p x SU(3)q}}

We now specialize to the case in which every simple factor is either $SU(2)$ or $SU(3)$.

 For each factor $G_i$ define
\begin{equation}
{\mathcal{Q}_i} :=
\sum_R
n_{i,R} c_{i,R}=c_{i,A}-\mathcal{C}_i.
\label{5.32}
\end{equation}
For the $SU(2)$ and $SU(3)$ factors considered in this section,
the coefficients $c_{R_i}$ are integers in our normalization.
Thus the quantities ${\mathcal{Q}_i}$ defined in Eq.~\eqref{5.32} are integers
already at the level of representation theory. In addition, the
integrality condition
\[
a\cdot b_i=\frac{{\mathcal{C}_i}}{6}\in \mathbb Z,
\]
imposes the congruence
\[
{\mathcal{Q}_i}\equiv c_{i,A}\pmod 6.
\]
Hence
\[
{\mathcal{Q}_i}\equiv 4 \pmod 6
\quad\text{for }SU(2),
\qquad
{\mathcal{Q}_i}\equiv 0 \pmod 6
\quad\text{for }SU(3).
\]

Then define 
\begin{equation}
\Delta_i :=
\sum_R
n_{i,R}
\left(
b_{i,R}
-\frac12 c_{i,R}
\right)
.
\label{5.33}
\end{equation}
one finds
\begin{equation}
{\mathcal{B}_i}
=
b_{i,A}
-
\Delta_i
-
\frac12 \mathcal{Q}_i .
\label{5.36}
\end{equation}

In the present analysis,
\begin{equation} 
\Delta_i \in 6 {\mathbb Z}_{\geq 0}, 
\end{equation}
because for any $SU(2)$ or $SU(3)$ irreducible representations the quantity
\begin{equation}
b_R-\frac12 c_R \in 6 {\mathbb Z}_{\geq 0}.
\end{equation}
In fact, for the $m$-dimensional irreducible representation of $SU(2)$,
\begin{equation}
b_m - \frac12 c_m
=
6\binom{m+2}{5}=\frac{1}{20} m (m^2-1) (m^2-4)\qquad m\geq 2,
\label{5.39}
\end{equation}
which is manifestly a nonnegative multiple of $6$. We quote the second expression, which also applies to $m=1$ and which gives a vanishing result. 
The same divisibility property also holds for $SU(3)$ representations. Indeed, let $(p,q)$ be the Dynkin labels of an irreducible representation of $SU(3)$ then 
\begin{equation}
b_{p,q}-\frac{1}{2}c_{p,q}=6 N_{p,q},
\end{equation}
where 
\begin{equation}
N_{p,q}
=
\frac{
(p+1)(q+1)(p+q+2)
\bigl(p^2+q^2+pq+3p+3q\bigr)
\bigl(p^2+q^2+pq+3p+3q-4\bigr)
}{720}\in {\mathbb Z}_{\geq 0}.
\end{equation}

This gives explicit expressions for $\Delta_i$.
For $SU(2)$ one obtains
\begin{equation}
\Delta_i
=
\frac{
({\mathcal{Q}_i}-6k+2)\big((k+1){\mathcal{Q}_i}-16k-88\big)
}{
12(k+3)^2
},
\label{5.40}
\end{equation}
while for $SU(3)$ one finds
\begin{equation}
\Delta_i
=
\frac{
({\mathcal{Q}_i}-6k+12)\big((k+1){\mathcal{Q}_i}-18k-114\big)
}{
12(k+3)^2
}.
\label{5.41}
\end{equation}

Thus every gauge factor must satisfy the arithmetic conditions
\begin{equation}
\Delta_i = \Delta_i({\mathcal{Q}_i};k),
\qquad
\Delta_i \in 6\mathbb Z_{\geq 0} .
\label{5.42}
\end{equation}
The simultaneous positivity and divisibility constraints on $\Delta_i$ are highly restrictive and are responsible for much of the observed rigidity.

The anomaly equations therefore split the search space into two disconnected arithmetic branches.
For $SU(2)$,
\begin{equation}
({\mathcal{Q}_i}-6k+2)\big((k+1){\mathcal{Q}_i}-16k-88\big)\ge 0,
\label{5.43}
\end{equation}
which implies either
\begin{equation}
{\mathcal{Q}_i}
\le
\min\left(
6k-2,
\frac{16k+88}{k+1}
\right),
\label{5.44}
\end{equation}
or
\begin{equation}
{\mathcal{Q}_i}
\ge
\max\left(
6k-2,
\frac{16k+88}{k+1}
\right).
\label{5.45}
\end{equation}
Similarly, for $SU(3)$ one obtains
\begin{equation}
({\mathcal{Q}_i}-6k+12)\big((k+1){\mathcal{Q}_i}-18k-114\big)\ge 0,
\label{5.46}
\end{equation}
so that either
\begin{equation}
{\mathcal{Q}_i}
\le
\min\left(
6k-12,
\frac{18k+114}{k+1}
\right),
\label{5.47}
\end{equation}
or
\begin{equation}
{\mathcal{Q}_i}
\ge
\max\left(
6k-12,
\frac{18k+114}{k+1}
\right).
\label{5.48}
\end{equation}

The additional integrality condition
\begin{equation}
\Delta_i \in 6\mathbb Z, 
\label{5.49}
\end{equation}
further discretizes the set of allowed values.

\subsection{Specialization to \texorpdfstring{$k=1$}{k=1}}

For $k=1$, we have 
\begin{equation}
\Delta_{SU(2)}
=
\frac{({\mathcal{Q}_i}-4)({\mathcal{Q}_i}-52)}{96},
\label{5.50}
\end{equation}
and
\begin{equation}
\Delta_{SU(3)}
=
\frac{({\mathcal{Q}_i}+6)({\mathcal{Q}_i}-66)}{96}.
\label{5.51}
\end{equation}. 

The first admissible values are
\begin{equation}
{\mathcal{Q}}_{SU(2)} \in \{4,52,76,100,124,148,172,196,\ldots\}.
\end{equation}
and
\begin{equation}
{\mathcal{Q}}_{SU(3)}
\in
\{66,90,114,138,162,186,210,234,258,\ldots\}.
\label{5.53}
\end{equation}

The values of different $\mathcal{Q}_i$ cannot be chosen independently. Once they are fixed, the vectors $b_i$ are determined, and the mixed anomaly equations
\begin{equation}
b_i \cdot b_j
=
\lambda_i\lambda_j C_{ij}
\label{5.54}
\end{equation}
must be realizable by actual matter contributions.

For example, the minimal choice
\begin{equation}
{\mathcal{Q}}_{SU(2)} = 4, \qquad {\mathcal{Q}}_{SU(3)} = 66
\end{equation}
gives
\begin{equation}
b_{SU(2)} = (-1,1), \qquad b_{SU(3)} = (1,4),
\end{equation}
and therefore
\begin{equation}
b_{SU(2)}\cdot b_{SU(3)} = -3.\label{eqn:first-branch}
\end{equation}
In fact, Eq.~\eqref{eqn:first-branch} holds for the first branch, where ${\mathcal{Q}}_{SU(2)} = 4,~{\mathcal{Q}}_{SU(3)} \geq 66$.

However, the mixed anomaly coefficients satisfy
\begin{equation}
 {\mathcal{C}_{ij}} = \sum_{RS} n_{ij,RS}c_{i, R }c_{j,S} ,\qquad  i \neq j.
\label{5.55}
\end{equation}
since all multiplicities, dimensions, and coefficients $c_{i,R}$ are
nonnegative. Because $\lambda_i=\lambda_j=1$ for both SU(2) and SU(3),
the mixed anomaly equation then implies
\begin{equation}
b_i\cdot b_j \ge 0.
\end{equation}
Consequently, this arithmetic branch is excluded immediately by the
mixed anomaly equations.

The two $k=1$ solutions found so far are:
\begin{itemize} 
\item {Model X}
\begin{equation}
{\mathcal{Q}}_{SU(2)} = 52,
\qquad
{\mathcal{Q}}_{SU(3)_1} = 258,
\qquad
{\mathcal{Q}}_{SU(3)_2} = 258 .
\label{5.59}
\end{equation}
\item{Model XI}
\begin{equation}
{\mathcal{Q}}_{SU(2)} = 196,
\qquad
{\mathcal{Q}}_{SU(3)_1} = 186,
\qquad
{\mathcal{Q}}_{SU(3)_2} = 186 .
\label{5.60}
\end{equation}
\end{itemize}
No additional solutions within the scanned representation domain were found after imposing the full set of arithmetic and mixed anomaly constraints.

In conclusion, the classification problem reduces to a highly constrained arithmetic search for each value of $k$.

The one-factor anomaly equations first reduce the search to a sparse arithmetic set of possible \({\mathcal{Q}}_i\). Within this reduced set, the mixed anomaly equations and the requirement that the resulting \({\mathcal{Q}}_i\), \(\Delta_i\), and \({\mathcal{C}}_{ij}\) be realized by actual product representations provide the decisive obstruction. In practice, most surviving arithmetic branches fail at this combined mixed-anomaly/representation-realizability stage. Although a complete classification theorem has not yet been established, the observed structure strongly suggests a remarkable degree of rigidity. In particular, the known solutions appear to occupy isolated arithmetic islands within the space of anomaly-free spectra.

A key advantage of this arithmetic analysis is that it provides an efficient new framework for constructing low-rank anomaly-free models. Without it, finding solutions that satisfy the Green–Schwarz factorization typically reduces to solving a system of Diophantine equations with a large number of variables, which is often intractable. In contrast, once a suitable group structure is fixed, this preliminary arithmetic analysis allows one to reformulate the problem in terms of algebraic equations for $k$, which is considerably more tractable in many cases.

\section{Discussion and outlook}
\label{sec:discussion-and-outlook}

In this paper we have presented several new anomaly-free six-dimensional $\mathcal{N}=(1,0)$ supergravity theories with gauged $U(1)_R$ symmetry. The models considered here satisfy the irreducible gravitational anomaly condition and admit Green--Schwarz factorization compatible with the structure of the anomaly polynomial. We have furthermore analyzed the associated unimodularity constraints on the dyonic string charge lattice  and investigated the arithmetic structure underlying the anomaly equations.

One of the central motivations for this work is the broader problem of classifying consistent six-dimensional supergravity theories. In theories without gauged R-symmetry, anomaly cancellation still permits a large number of solutions, and the resulting landscape of anomaly-free models appears comparatively abundant. By contrast, the gauged $U(1)_R$ theories studied here exhibit a striking degree of rigidity. The combined constraints arising from anomaly cancellation, Green--Schwarz factorization, global consistency conditions, and charge-lattice structure drastically reduce the number of allowed theories. This suggests that six-dimensional supergravities with gauged $U(1)_R$ symmetry may be substantially more constrained than their ungauged counterparts.

Several structural features repeatedly emerged in our analysis. First, all currently known examples appear to possess unimodular charge lattices. In the present work we showed that for theories with $n_T=1$ and gauge group of the form
\begin{equation}
G=G_\mathrm{non-Abelian} \times U(1)_R,
\end{equation}
the condition
\begin{equation}
n_V \equiv 8 \pmod{12}
\end{equation}
is necessary and sufficient for embedding the anomaly coefficients into a unimodular lattice.

A second recurring feature concerns the type of the string charge lattice. The $L_{3,\mathrm{even}}$ lattice never appears once the congruence condition is applied. It is suggestive that this type of lattice may be excluded using representation theory analysis at the spectra level.

The analysis of theories with gauge group
\begin{equation}
SU(2)^p \times SU(3)^q \times U(1)_R
\end{equation}
also revealed a remarkable arithmetic structure. The anomaly equations reduce the classification problem to a finite search governed by a small number of integer parameters, and the mixed anomaly constraints eliminate the overwhelming majority of naively allowed solutions. The resulting structure suggests a high degree of rigidity and hints at the possibility of a complete classification theorem for at least certain subclasses of gauged $U(1)_R$ theories.

Another natural direction concerns the relation between anomaly-free supergravity theories and ultraviolet completion in string theory. Many researches have been conducted on string/M/F-theory realization of ungauged or non-$R$ gauged 6D $\mathcal{N}=(1,0)$ supergravity. In those cases, anomaly-free 6D models can often be lifted to a higher dimension. For example, in \cite{Grimm:2013fua}, the authors considered F-theory realization of models where vector fields gauged axionic shift symmetries in the hypermultiplet sector. A conjecture about string universality in six dimensions was further proposed in \cite{Kumar:2009us}. Nevertheless, $R$ symmetry gauged 6D supergravity still lacks good realizations in string or F-theory compactifications\footnote{A non-compact compactification of $U(1)_R$ gauged supergravity was discussed in \cite{Duff:1996cf}. This construction, however, does not provide a string realization of an anomaly-free six-dimensional theory, since the ten-dimensional Yang–Mills sector and the additional matter required for anomaly cancellation were not included.}. It would therefore be very interesting to determine whether any anomaly-free gauged $U(1)_R$ supergravities listed above admit ultraviolet completions, or whether some belong to a larger set of apparently consistent low-energy theories. It should also be interesting to apply additional constraints like the ones proposed in \cite{Kim:2019vuc,Kim:2024tdh}.

An important related problem concerns six-dimensional supergravity theories with gauged $Sp(1)_R$ symmetry. At present, no anomaly-free examples of such theories are known. The methods developed in the present work, including the analysis of anomaly factorization, unimodularity constraints, and arithmetic restrictions on the matter content, provide a natural starting point for a systematic investigation of $Sp(1)_R$-gauged models. We hope to return to this problem in future work.

\section*{Acknowledgements}
We thank X. Guo and E. Sezgin for useful discussions. This work is supported by the NSF grant PHY-2413006 and endowment funds from the Mitchell Family Foundation.
\appendix

\section{Group-theoretical coefficients}
\label{app:group-coeff}

\begin{table}[htbp]
    \scriptsize
    \centering
    \begin{tabular}{|c|c|c|c|c|c|}
    \toprule
     Group & Irrep $ R $ & $a_{ R }$ & $b_{ R }$ & $c_{ R }$ & Reality \\
    \midrule
    \multirow{3}{*}{$SU(2)$}
    & $\bf{2}$ & $0$ & $\frac{1}{2}$ & $1$ & pseudo-real \\
    & $\bf{3}$ & $0$ & $8$ & $4$ & real \\
    & $\bf{4}$ & $0$ & $41$ & $10$ & pseudo-real \\
    \midrule
    \multirow{4}{*}{$SU(3)$}
    & $\bf{3}$ & $0$ & $\frac{1}{2}$ & $1$ & complex \\
    & $\bf{8}$ & $0$ & $9$ & $6$ & real\\
    & $\bf{10}$ & $0$ & $\frac{99}{2}$ & $15$ &complex \\
    & $\bf{15}$ & $0$ & $\frac{371}{2}$ & $35$ & complex \\ 
    \midrule
    \multirow{6}{*}{$SU(6)$}
    & $\bf{6}$ & $1$ & $0$ & $1$ & complex\\
    & $\bf{15}$ & $-2$ & $3$ & $4$ & complex\\
    & $\bf{20}$ & $-6$ & $6$ & $6$ & pseudo-real\\
    & $\bf{21}$ & $14$ & $3$ & $8$ & complex \\
    & $\bf{35}$ & $12$ & $6$ & $12$ & real\\
    & $\bf{540}$ & $18$ & $684$ & $378$ & pseudo-real\\
    \midrule
    \multirow{3}{*}{$SU(18)$}
    & $\bf{18}$ & $1$ & $0$ & $1$ & complex\\
    & $\bf{323}$ & $36$ & $6$ & $36$ & real\\
    & $\bf{816}$ & $36$ & $42$ & $120$ & complex\\
    \midrule
    \multirow{3}{*}{$Spin(7)$}
    & $\bf{7}$ & $1$ & $0$ & $1$ & real\\
    & $\bf{8}$ & $-\frac{1}{2}$ & $\frac{3}{8}$ & $1$ & real\\
    & $\bf{21}$ & $-1$ & $3$ & $5$ & real\\
    \midrule
    \multirow{2}{*}{$Spin(10)$}
    & $\bf{10}$ & $1$ & $0$ & $1$ & real\\
    & $\bf{45}$ & $2$ & $3$ & $8$ & real\\
    \midrule
    \multirow{3}{*}{$Spin(12)$}
    & $\bf{12}$ & $1$ & $0$ & $1$ & real\\
    & $\bf{32}$ & $-2$ & $\frac{3}{2}$ & $4$ & pseudo-real \\
    & $\bf{66}$ & $4$ & $3$ & $10$ & real\\
    \midrule
    \multirow{3}{*}{$Spin(14)$}
    & $\bf{14}$ & $1$ & $0$ & $1$ & real\\
    & $\bf{91}$ & $6$ & $3$ & $12$ & real\\
    & $\bf{364}$ & $6$ & $30$ & $66$ & real\\
    \midrule
    \multirow{3}{*}{$Spin(15)$}
    & $\bf{15}$ & $1$ & $0$ & $1$ & real\\
    & $\bf{105}$ & $7$ & $3$ & $13$ & real\\
    & $\bf{128}$ & $-8$ & $6$ & $16$ & real\\
    \midrule
    \multirow{3}{*}{$Sp(2)$}
    & $\bf{4}$ & $1$ & $0$ & $1$ & pseudo-real \\
    & $\bf{5}$ & $-4$ & $3$ & $2$ & real\\
    & $\bf{10}$ & $12$ & $3$ & $6$ & real\\
    \midrule
    \multirow{5}{*}{$Sp(5)$}
    & $\bf{10}$ & $1$ & $0$ & $1$ & pseudo-real \\
    & $\bf{55}$ & $18$ & $3$ & $12$ & real\\
    & $\bf{110}$ & $-9$ & $18$ & $27$ & pseudo-real \\
    & $\bf{132}$ & $-42$ & $42$ & $42$ & pseudo-real \\
    & $\bf{220}$ & $162$ & $42$ & $78$ & pseudo-real \\
    \midrule
    \multirow{3}{*}{$Sp(9)$}
    & $\bf{18}$ & $1$ & $0$ & $1$ & pseudo-real \\
    & $\bf{152}$ & $10$ & $3$ & $16$ & real\\
    & $\bf{171}$ & $26$ & $3$ & $20$ & real\\
    \midrule
    \multirow{2}{*}{$F_4$}
    & $\bf{26}$ & $0$ & $\frac{1}{12}$ & $1$ & real\\
    & $\bf{52}$ & $0$ & $\frac{5}{12}$ & $3$ & real\\
    \midrule
    \multirow{2}{*}{$E_6$}
    & $\bf{27}$ & $0$ & $\frac{1}{12}$ & $1$ & complex\\
    & $\bf{78}$ & $0$ & $\frac{1}{2}$ & $4$ & real\\
    \midrule
    \multirow{2}{*}{$E_7$}
    & $\bf{56}$ & $0$ & $\frac{1}{24}$ & $1$ & pseudo-real \\
    & $\bf{133}$ & $0$ & $\frac{1}{6}$ & $3$ & real\\
    \midrule
    \multirow{1}{*}{$E_8$}
    & $\bf{248}$ & $0$ & $\frac{1}{100}$ & $1$ & real\\
    \bottomrule
    \end{tabular}
    \caption{Group-theoretical coefficients relevant to the models in this paper.}
    \label{tab:group-coefficients}
\end{table}

We follow the conventions and notations in \cite{Avramis:2005hc} to list the group-theoretical coefficients, namely $a_R$, $b_R$, and $c_R$, for Lie groups/algebras relevant to \autoref{sec:new-models}. For convenience, we restate some formulae here.
\begin{equation}
    c_ R =\frac{l_2( R )}{l_2(\mathcal{F})},
\end{equation}
\begin{equation}
    b_ R =\frac{3}{2+\mathrm{dim}(\mathcal{A})}[\frac{l_2( R )}{l_2(\mathcal{F})}]^2[\frac{\mathrm{dim}(\mathcal{A})}{\mathrm{dim}( R )}-\frac{1}{6}\frac{l_2(\mathcal{A})}{l_2( R )}],
\end{equation}
\begin{equation}
    a_ R =l_4( R ),
\end{equation}
where $l_2( R )$ and $l_4( R )$ are respectively the second and fourth order Dynkin indices of the representation $ R $, $\mathcal{F}$ denotes the fundamental representation and $\mathcal{A}$ denotes the adjoint.\footnote{The normalization of the $l_4( R )$ is chosen such that $l_4(\mathcal{F})=1$, while the normalization for $l_2( R )$ is not important to us.} These coefficients are listed in \autoref{tab:group-coefficients}, where the reality of representations can be found in \cite{Slansky:1981yr}.

\section{Fourth-order Casimir for $D_4$}
\label{app:4th-D4}

In this appendix, we derive the full conditions for the vanishing of fourth-order invariants for $D_4$ algebra, following \cite{Okubo:1981td}. There the author classified all simple Lie algebras into three types based on the dimension of $V$, the space of all fourth-order Casimir invariants. Type I includes $A_1$, $A_2$ and all exceptional Lie algebras, which do not have a genuine fourth-order invariant, meaning that all are proportional to the square of the unique second-order invariant(up to a constant) and $\mathrm{dim}V=1$. Type II includes all other simple Lie algebras except $D_4$. This type of Lie algebras has one genuine fourth-order Casimir, so for them $\mathrm{dim}V=2$. Type III Lie algebra is the most special one and only includes $D_4$, whose $\mathrm{dim}V=3$. 

Let $h_{\mu\nu\alpha\beta}$ and $H_{\mu\nu\alpha\beta}( R )$ denote the totally symmetrized tensor in the reference representation (which we choose to be the fundamental) and in an arbitrary irreducible representation,
\begin{equation}
    h_{\mu\nu\alpha\beta}=\frac{1}{4!}\sum_{\mathrm{sym.}}\mathrm{tr}(T_\mu T_\nu T_\alpha T_\beta),
\end{equation}
\begin{equation}
    H_{\mu\nu\alpha\beta}=\frac{1}{4!}\sum_{\mathrm{sym.}}\mathrm{Tr}(T_\mu T_\nu T_\alpha T_\beta),
\end{equation}
where $T_\mu$ denotes generators of $D_4$. For later purposes, also define $g_{\mu\nu\alpha\beta}$ and $G_{\mu\nu\alpha\beta}$ as
\begin{equation}
    g_{\mu\nu\alpha\beta}=[2+\mathrm{dim}(\mathcal{A})]h_{\mu\nu\alpha\beta}-\frac{H(\mathcal{F})}{I_2(\mathcal{F})}\{g_{\mu\nu}g_{\alpha\beta}+g_{\mu\alpha}g_{\nu\beta}+g_{\mu\beta}g_{\nu\alpha}\},
\end{equation}
\begin{equation}
\begin{aligned}
    G_{\mu\nu\alpha\beta}&=\frac{1}{4!}\sum_{\mathrm{sym.}}\mathrm{Tr}(T_\mu T_\nu T_\alpha T_\beta)-\frac{1}{3}K( R )\{[\mathrm{Tr}(T_\mu T_\nu)][\mathrm{Tr}(T_\alpha T_\beta]\\&+[\mathrm{Tr}(T_\mu T_\alpha)][\mathrm{Tr}(T_\nu T_\beta]+[\mathrm{Tr}(T_\mu T_\beta)][\mathrm{Tr}(T_\nu T_\alpha] \}\\
    &=H_{\mu\nu\alpha\beta}-\frac{1}{2+\mathrm{dim}(\mathcal{A})}\frac{\mathrm{dim}( R )H( R )}{\mathrm{dim}(\mathcal{F})I_2(\mathcal{F})}\{g_{\mu\nu}g_{\alpha\beta}+g_{\mu\alpha}g_{\nu\beta}+g_{\mu\beta}g_{\nu\alpha}\},
\end{aligned}  
\end{equation}
where $g$ is the Killing form, $H( R )=[I_2( R )]^2-\frac{1}{6}I_2(\mathcal{A})I_2( R )$, $I_2( R )$ is the second-order Casimir whose normalization is not important to us (in fact, only their ratio matters), and $K( R )=\frac{\mathrm{dim}(\mathcal{A})}{2[2+\mathrm{dim}(\mathcal{A})]\mathrm{dim}( R )}[6-\frac{I_2(\mathcal{A})}{I_2( R )}]$. One can denote the quadratic trace terms in $G_{\mu\nu\alpha\beta}$ by $A_{\mu\nu\alpha\beta}$, and use $\mathrm{dim}(\mathcal{F})=8$ and $\mathrm{dim}(\mathcal{A})=28$ to rewrite
\begin{equation}
    g_{\mu\nu\alpha\beta}=30h_{\mu\nu\alpha\beta}-\frac{240}{\mathrm{dim}( R )}\frac{6-\phi(\mathcal{A})}{6\phi^2( R )-\phi(\mathcal{A})\phi( R )}A_{\mu\nu\alpha\beta},
\end{equation}
where $\phi( R )=\frac{I_2( R )}{I_2(\mathcal{F})}$.

For type I simple Lie algebras that do not possess a genuine fourth-order Casimir invariants, $g_{\mu\nu\alpha\beta}=G_{\mu\nu\alpha\beta}=0$. For type II Lie algebras, they satisfy Eq.~(2.23) in \cite{Okubo:1981td}
\begin{equation}
    [2+\mathrm{dim}( R )]G_{\mu\nu\alpha\beta}=B( R )g_{\mu\nu\alpha\beta},\label{eq：G-proportion-to-g}
\end{equation}
where $B( R )=\frac{\mathrm{dim}( R )J_4( R )}{\mathrm{dim}(\mathcal{F})J_4(\mathcal{F})}$, with $J_4( R )$ being the eigenvalue of the modified fourth-order Casimir invariant $J_4=g^{\mu\nu\alpha\beta}T_\mu T_\nu T_\alpha T_\beta$, can be explicitly computed using formulae in section 3 of \cite{Okubo:1981td}.

For $D_4$ algebra, since there exists another independent fourth-order Casimir, Eq.~\eqref{eq：G-proportion-to-g} is modified to be
\begin{equation}
        [2+\mathrm{dim}( R )]G_{\mu\nu\alpha\beta}=B( R )g_{\mu\nu\alpha\beta}+C( R )e_{\mu\nu\alpha\beta}.\label{eq:G-g-e}
\end{equation}
In the above equation, $e_{\mu\nu\alpha\beta}=\epsilon_{a_1b_1...a_4b_4}$, and $C( R )$ can be computed using Eq.~(4.12b) in \cite{Okubo:1981td} to be
\begin{equation}
    C( R )=\frac{2\mathrm{dim}( R )\hat{I}_4( R )}{7},
\end{equation}
where $\epsilon_{a_1b_1...a_4b_4}$ is the Levi-Civita symbol in eight-dimensional space (here one may relabel $\mu_i$ as the anti-symmetric pairs $(a_i,b_i),~a_i<b_i, i=1,2,3,4$) and $\hat{I}_4( R )$ is a representation-dependent constant defined in Eq.~(4.8) in \cite{Okubo:1981td}.

Now we are ready to express $\mathrm{Tr}_R F^4$ in quartic/quadratic trace terms in the fundamental representation for $D_4$. Let $F=F^\mu T_\mu$ denote an arbitrary $D_4$ element in any irreducible representation. Contracting Eq.~\eqref{eq:G-g-e} with $F^\mu F^\nu F^\alpha F^\beta$ and making use of $\mathrm{dim}(\mathcal{F})=8$ and $\mathrm{dim}(\mathcal{A})=28$, one gets
\begin{align}
    30\{\mathrm{Tr}_ R(F^4)-K( R )[\mathrm{Tr}_ R(F^2)]^2\}=&B( R )\{{30}\mathrm{tr}(F^4)-\frac{240}{\mathrm{dim}( R )}\frac{6-\phi(\mathcal{A})}{6\phi^2( R )-\phi(\mathcal{A})\phi( R )}\\
    &K( R )[\mathrm{Tr}_ R(F^2)]^2\}+\frac{2\mathrm{dim}( R )\hat{I}_4( R )}{7}\tilde{F}^4,
\end{align}
where $\tilde{F}^4=e_{\mu\nu\alpha\beta}F^\mu F^\nu F^\alpha F^\beta$. After some algebra, we obtain
\begin{equation}
    \mathrm{Tr}_ R(F^4)=B( R )\mathrm{tr}(F^4)+g( R )K( R )c_ R ^2[\mathrm{tr}(F^2)]^2+\frac{\mathrm{dim}( R )\hat{I}_4( R )}{105}\tilde{F}^4,
\end{equation}
where $g( R )=1-\frac{8B( R )}{\mathrm{dim}( R )}\frac{6-\phi(\mathcal{A})}{6\phi^2( R )-\phi(\mathcal{A})\phi( R )}$. One can then read off
\begin{equation}
    \begin{aligned}
    a_ R =B( R ),~~~b_ R =g( R )K( R )c_ R ^2,~~~d_ R =\frac{\mathrm{dim}( R )\hat{I}_4( R )}{105}=\frac{C( R )}{30}.
    \end{aligned}
\end{equation}
Because $\tilde{F}^4$ in general cannot be factorized into $[\mathrm{tr}(F^2)]^2$, one has to make sure that its coefficient cancels when adding up relevant terms in the anomaly polynomial for all hyperfermions. In fact, it turns out that as long as there are equivalent numbers of hypermultiplets transforming under a pair of chiral representations, such as $\bf{8_s}$ and $\bf{8_c}$, this coefficient of $\tilde{F}^4$ will vanish automatically since their $\hat{I}_4( R )$ differ by a sign\footnote{In fact, this is a sufficient but not necessary condition because one cannot rule out the possibility that, for example, hyperfermions that transform under $\bf{8_s}$ and $\bf{56_c}$ will not conspire to cancel the $\tilde{F}^4$ term as long as the numbers of multiplets match perfectly.}.

\end{document}